\documentclass[reprint,twocolumn,aps,superscriptaddress]{revtex4-2}
\usepackage{amsmath,amssymb,amsfonts,mathrsfs}
\usepackage{bm}
\usepackage{graphicx}
\usepackage{hyperref}
\usepackage{amsbsy}
\usepackage{dcolumn}
\usepackage{lipsum}

\hypersetup{
	colorlinks = true,
	citecolor  = red,  
	linkcolor  = blue  
}

\begin{document}
	
	
	\title{Parametric phase modulation in superconducting circuits} 
	
	
	\author{Zhuang Ma}
	\thanks{These three authors contributed equally to this work.}
	\affiliation{National Laboratory of Solid State Microstructures, School of Physics, Nanjing University, Nanjing 210093, China}
	\affiliation{Shishan Laboratory, Suzhou Campus of Nanjing University, Suzhou 215000, China}
        \affiliation{Jiangsu Key Laboratory of Quantum Information Science and Technology, Nanjing University, Suzhou 215163, China}
	\author{Xianke Li}
	\thanks{These three authors contributed equally to this work.}
	\affiliation{National Laboratory of Solid State Microstructures, School of Physics, Nanjing University, Nanjing 210093, China}
	\affiliation{Shishan Laboratory, Suzhou Campus of Nanjing University, Suzhou 215000, China}
        \affiliation{Jiangsu Key Laboratory of Quantum Information Science and Technology, Nanjing University, Suzhou 215163, China}
	\author{Hongyi Shi}
	\thanks{These three authors contributed equally to this work.}
	\affiliation{National Laboratory of Solid State Microstructures, School of Physics, Nanjing University, Nanjing 210093, China}
	\affiliation{Shishan Laboratory, Suzhou Campus of Nanjing University, Suzhou 215000, China}
        \affiliation{Jiangsu Key Laboratory of Quantum Information Science and Technology, Nanjing University, Suzhou 215163, China}
	\author{Ruonan Guo}
	\affiliation{National Laboratory of Solid State Microstructures, School of Physics, Nanjing University, Nanjing 210093, China}
	\affiliation{Shishan Laboratory, Suzhou Campus of Nanjing University, Suzhou 215000, China}
        \affiliation{Jiangsu Key Laboratory of Quantum Information Science and Technology, Nanjing University, Suzhou 215163, China}
	\author{Jianwen Xu}
	\affiliation{National Laboratory of Solid State Microstructures, School of Physics, Nanjing University, Nanjing 210093, China}
	\affiliation{Shishan Laboratory, Suzhou Campus of Nanjing University, Suzhou 215000, China}
        \affiliation{Jiangsu Key Laboratory of Quantum Information Science and Technology, Nanjing University, Suzhou 215163, China}
	\author{Xinsheng Tan}
	\email{tanxs@nju.edu.cn}
	\affiliation{National Laboratory of Solid State Microstructures, School of Physics, Nanjing University, Nanjing 210093, China}
	\affiliation{Shishan Laboratory, Suzhou Campus of Nanjing University, Suzhou 215000, China}
        \affiliation{Jiangsu Key Laboratory of Quantum Information Science and Technology, Nanjing University, Suzhou 215163, China}
	\affiliation{Synergetic Innovation Center of Quantum Information and Quantum Physics, University of Science and Technology of China, Hefei, Anhui 230026, China}
	\affiliation{Hefei National Laboratory, Hefei 230088, China}

	\author{Yang Yu}
	\affiliation{National Laboratory of Solid State Microstructures, School of Physics, Nanjing University, Nanjing 210093, China}
	\affiliation{Shishan Laboratory, Suzhou Campus of Nanjing University, Suzhou 215000, China}
        \affiliation{Jiangsu Key Laboratory of Quantum Information Science and Technology, Nanjing University, Suzhou 215163, China}
	\affiliation{Synergetic Innovation Center of Quantum Information and Quantum Physics, University of Science and Technology of China, Hefei, Anhui 230026, China}
	\affiliation{Hefei National Laboratory, Hefei 230088, China}
 
	
	\date{\today}
	
	\begin{abstract}
	Parametric modulation, valued for its versatility, is widely employed in superconducting circuits for quantum simulations and high-fidelity two-qubit gates. Conventionally, the qubit coupling strength is determined by the amplitude of the parametric flux pulse, which affects qubit parameters dramatically. In this article, we propose and implement a phase-modulation scheme to tune the interaction strength via adjustment of the relative phase between the parametric flux pulses applied to two coupled qubits. We characterize this modulation for sideband couplings, at both sweet and off-sweet spots, achieving a broad range of coupling strengths, as confirmed by both population dynamics and spectroscopy methods. This approach enables phase-controlled modulation of coupling strength, providing a promising  candidate for parametrically driven quantum simulations and gate operations.
	\end{abstract}
	
	
	\maketitle 
	

	\section{Introduction} 
	Superconducting quantum circuits have emerged as a leading platform for large-scale quantum simulation and computation due to their high controllability, flexibility, and scalability \cite{Krantz2019}. To enable the scaling of superconducting qubits, various tunable coupling schemes have been proposed to address the challenges of large-scale systems, where couplings between qubits must be precisely tuned to suppress unwanted interactions or enhance desired ones \cite{Niskanen2007, Chen2014, Yan2018}. 
	
	To meet this demand, various schemes for tunable coupling have been developed. Tunable couplers, often realized as additional circuit elements, are widely implemented in transmon-based architectures to mediate qubit-qubit interactions, allowing couplings to be turned on or off, and mitigating issues like parasitic coupling, frequency crowding, control crosstalk, and leakage to non-computational states \cite{Yan2018, Sete2021a}. These advantages have facilitated the realization of large-scale quantum simulations and high-fidelity two-qubit gates \cite{Arute2019, Wu2021}.
	
	 Complementary to these hardware-based couplers, some tunable coupling schemes exploit the inherent properties of superconducting qubits, such as parametric modulation \cite{McKay2016, Didier2018, Caldwell2018} and all-microwave schemes \cite{Rigetti2010, Krinner2020}. Two-qubit gates activated by parametric modulation have gained increasing attention due to their robustness against flux distortions, noise, and crosstalk \cite{Reagor2018, Caldwell2018, Fried2019, Chu2020, Han2020, Ganzhorn2020, Abrams2020, Sete2021, Li2022, Sete2024}. Periodic modulation provides net-zero and refocusing effects \cite{Rol2019, Ma2024}, which lead to dynamical sweet spots \cite{Didier2019, Didier2019a, Hong2020, Mundada2020, Huang2021, Valery2022, BrisenoColunga2025}, the continuous version of the dynamical decoupling scheme \cite{Ma2024}, and enhanced qubit coherence \cite{Matityahu2021, Matityahu2024, Wudarski2024} (see Appendix~\ref{app:coherence} for details). Parametric pulses can bridge the energy gaps between the far-detuned qubit states, inducing sideband interactions that help avoid frequency collisions, thus offering great flexibility for quantum simulation and computation \cite{Chu2020, Li2021, Sete2021, Zheng2022, Liu2023, Ma2023, Zhang2024}. This modulation enables transitions between off-resonance qubits with tunable coupling, even for parametric-resonance qubits \cite{Sete2021}. 

     In conventional parametric modulation schemes, the coupling strength is governed by the amplitude and frequency of the applied pulses. However, tuning these parameters typically induces substantial shifts in the qubit frequencies, thereby imposing significant challenges on the calibration and scalability of large-scale quantum processor architectures.
     To overcome this limitation, we introduce an approach to control the interaction strength via adjustment of the phase of the parametric modulation. Normally, the parametric phase is utilized in specific quantum simulation protocols for tailoring Hamiltonians \cite{Li2018, Cai2019, Zheng2022, Zhao2022, Rosen2024, Jiang2025} or for dynamic on-off switching of couplings under particular resonance conditions \cite{Wang2019, Liu2020, Shi2023}. Achieving precise phase control and compensation remains a significant challenge in both quantum gate operations \cite{Chu2020, Sete2021, Ma2023} and simulations \cite{Zheng2022, Zhang2024} based on parametric modulation.
  
    In this work, we re-examine the conventional role of the parametric phase, treating it not as an experimental overhead requiring compensation, but as a valuable and versatile quantum control resource. We propose and experimentally demonstrate a new approach where the relative phase between two simultaneous parametric flux pulses---one applied to each of two interacting superconducting transmon qubits---provides a direct control parameter for their coupling strength. A crucial advantage of this phase-based method is that it allows for precise modulation of the interaction strength without inducing the time-averaged qubit frequency shifts that typically necessitate cumbersome recalibration.

    We experimentally validate this mechanism in a versatile system of two qubits coupled via a tunable coupler. This configuration allows us to retain the inherent advantages of couplers while simultaneously enhancing the generality and flexibility of our parametric phase-modulation scheme. Our experiments, employing population dynamics and spectroscopy methods, systematically characterize this phase-controlled interaction at both sweet and off-sweet spots, thereby demonstrating its broad applicability under diverse operating conditions.

	\section{Theory}
	\subsection{Parametric modulation}
	In practical frequency-tunable transmon qubits, the relationship between the qubit frequency and externally applied flux biases of qubits is nonlinear. An arbitrary waveform generator (AWG) generates a programmed voltage pulse, $V(t) = \bar{V}+\tilde{V}\cos(\omega_pt+\phi_p)$, which is applied to the qubit, and the modulation of the flux bias pulse is described as
	\begin{equation}
		\Phi(t)=\bar{\Phi}+\tilde{\Phi}\cos(\omega_p t+\phi_p),
	\end{equation}
	where the flux oscillates around the parking flux bias $\bar{\Phi}$ with parametric amplitude $ \tilde{\Phi}$, frequency $\omega_p$, and phase $\phi_p$. The qubit frequency becomes time-dependent and can be expressed as the Fourier series
	\begin{equation}
		\omega(t)=\sum_{k=0}^{\infty} f_{k} \cos \left[k\left(\omega_{p} t+\phi_{p}\right)\right],
	\end{equation}
	where the Fourier coefficients are given by $$ f_k=\frac{\omega_p}{\pi(1+\delta_{k,0})}\int_0^{{2\pi}/{\omega_p}}dt\cos[k(\omega_pt+\phi_p)]\omega(t)$$. The frequency can be approximated as:
	\begin{equation}
		\omega(t) \approx \bar{\omega} + \epsilon_p \cos(2\omega_{p}t + 2\phi_{p}),
	\end{equation}
	when the parking flux bias $\bar{\Phi}$ is set at the sweet spot of the qubit and $ \tilde{\Phi} \lesssim \Phi_0/2$ (with $\Phi_0=h/2e$ being the magnetic flux quantum). Here, $\bar{\omega}$ denotes the time-averaged qubit frequency under modulation, while $\epsilon_p$ represents the qubit frequency excursion (i.e., the amplitude of the frequency modulation). If the parking flux bias $\bar{\Phi}$ is set at the off-sweet spot, the frequency can be approximated as
	\begin{equation}
		\omega(t) \approx \bar{\omega} + \epsilon_p \cos(\omega_{p}t + \phi_{p}),
	\end{equation} 
	with a very weak parametric amplitude $ \tilde{\Phi}$ (linear approximation).

 
	We illustrate our scheme using two qubits with a fixed coupling strength, described by the Duffing-oscillator Hamiltonian in the energy basis (with $\hbar=1$ assumed hereafter)
	\begin{equation}
		\begin{split}
			\mathcal{H}_0 &= \omega_1 b_1^{\dagger} b_1 + \frac{\alpha_1}{2} b_1^{\dagger} b_1^{\dagger} b_1b_1 +  \omega_2 b_2^{\dagger} b_2 + \frac{\alpha_2}{2} b_2^{\dagger} b_2^{\dagger} b_2b_2,\\
			\mathcal{H}_I  &=g (b_1 + b_1^{\dagger} ) (b_2 + b_2^{\dagger} ),
			\label{hamiltonian_lab}
		\end{split}
	\end{equation}
	where $ \omega_{1,2}, ~\alpha_{1,2},~b_{1,2}$,  $b^\dagger_{1,2}$, and $g$ represent the frequencies, anharmonicities, annihilation and creation operators, and bare coupling strength of qubits, respectively. When the parking flux bias $\bar{\Phi}_1$ is set at the off-sweet spot with a weak parametric amplitude $ \tilde{\Phi}_1$, modulating $Q_1$ as $$\omega_{1}(t) \approx \bar{\omega}_{1}+\epsilon_{p1} \cos(\omega_{p1} t + \phi_{p1})$$, it induces frequency-modulation sidebands, and the qubit frequency oscillates at many harmonics of the parametric frequency $\omega_{p1}$ (see Appendixes~\ref{app:spectroscopy} and \ref{app:taylor} for details). In principle, all parameters $ \omega_{1,2}, ~\alpha_{1,2}$, and $g$ under modulation in the frequency domain become time-dependent due to the interaction \cite{Didier2018, Sete2021, Perez2023}. 
	
	To obtain the effective Hamiltonian in a rotating frame, we define the unitary rotation transformation as $$U = \exp\left(-i \int_{0}^{t} \mathcal{H}_0(\tau) d \tau \right).$$
	For simplicity, we also define $F_1 (t) = \int_0^t \omega_1(\tau)d\tau$,$ F_2 (t) = \int_0^t \omega_2(\tau)d\tau$, $A_1 (t) = \int_0^t \alpha_1(\tau)d\tau$, and $A_2 (t) = \int_0^t \alpha_2(\tau)d\tau$. The effective Hamiltonian is then given by
	\begin{widetext}
			\begin{equation}
			\begin{split}
				H_{\mathrm{eff}} = i \frac{d U^{\dagger}}{d t} U+U^{\dagger} (\mathcal{H}_0+\mathcal{H}_I) U
				= &g(b_1b_2 \exp\{i[-F_1-A_1 (b_1^{\dagger} b_1-I) -F_2-A_2(b_2^{\dagger} b_2-I)]\} \\
				&+b_1^\dagger b_2 \exp\{i[F_1+A_1 b_1^{\dagger} b_1 -F_2-A_2(b_2^{\dagger} b_2-I)]\}\\
				&+b_1 b_2^\dagger \exp\{i[-F_1-A_1 (b_1^{\dagger} b_1-I) +F_2+A_2b_2^{\dagger} b_2]\}\\
				&+b_1^\dagger b_2^\dagger \exp\{i(F_1+A_1 b_1^{\dagger} b_1 + F_2+A_2b_2^{\dagger} b_2)\}),
				\label{hamiltonian_rot}
			\end{split}
		\end{equation}
	\end{widetext}
	where we approximate $ g $ as constant. In the $\{ |01\rangle, | 10\rangle \}$ subspace (indexing the states of coupled qubits $|Q_1 Q_2\rangle$), the above effective Hamiltonian in Eq. \eqref{hamiltonian_rot} is truncated as 
\begin{widetext}
	\begin{equation}
		\begin{split}
			H_{\mathrm{eff}} &= g\exp{i\Delta t}  \exp\left({i\frac{\epsilon_{p1}}{\omega_{p1}}[\sin(\phi_{p1})-\sin(\omega_{p1}t+\phi_{p1}) ]}\right)|10\rangle \langle 01| + \text{H.c.}\\
			&= g\sum_{n=-\infty}^{\infty}  J_{n}\left(\frac{\epsilon_{p1}}{\omega_{p1}}\right)\exp{i[(\Delta +n \omega_{p1})t+\beta_n]}|10\rangle \langle 01| + \text{H.c.},
			\label{coupling}
		\end{split}
	\end{equation}
\end{widetext}
	where we use the Jacobi-Anger expansion $e^{i z \sin \theta}=\sum_{n=-\infty}^{\infty}  J_{n}(z) e^{i n \theta}$ (with $J_{n}$ being the $n$-th Bessel function of the first kind), $\Delta = \bar{ \omega}_2 - \bar{\omega}_1$, and $\beta_n =  n(\phi_{p1}+\pi) + \frac{\epsilon_{p1}}{\omega_{p1}}\sin(\phi_{p1})$. The induced effective coupling strength is given by
	\begin{equation}
		g_{\text{eff}}^n = g  J_{n} \left(\frac{\epsilon_{p1}}{\omega_{p1}}\right),
		\label{effcoupling}
	\end{equation}
	when the resonance condition $\Delta + n\omega_{p1} = 0, n \in \mathbb{Z},$ is satisfied.

	\subsection{Parametric phase modulation}
	Here, we observe that the parametric phase $\phi_{p1}$ does not affect the effective coupling strength $g_\mathrm{eff}^n$ in Eq. \eqref{effcoupling}. However, introducing an additional modulation at $Q_2$ with a weak amplitude $\tilde{\Phi}_2$, given by $$\omega_{2}(t) \approx \bar{\omega}_{2}+\varepsilon_{p2} \cos(\omega_{p2} t + \phi_{p2}),$$ renders the relative phase $\delta \phi_p = \phi_{p1}-\phi_{p2}$ as a modulating factor for the effective coupling strength. According to  Eq. \eqref{hamiltonian_rot} and \eqref{coupling}, the new phase-tunable coupling strength is 
	
\begin{widetext}
	\begin{equation}
		\begin{split}
			g_{\text{phase}}^n 
			&=C_{\phi}\exp{(i\Delta t)}  \exp\left\{i\left[\frac{\epsilon_{p2}}{\omega_{p2}}\sin(\omega_{p2}t+\phi_{p2})-\frac{\epsilon_{p1}}{\omega_{p1}}\sin(\omega_{p1}t+\phi_{p1})\right]\right\}\\
			&=C_{\phi}\exp{(i\Delta t)} \exp\left[i  A\sin\left(\frac{\omega_{p1}+\omega_{p2}}{2}t+\frac{\phi_{p1}+\pi + \phi_{p2}}{2} + \varphi\right)\right]	\\
			&= C_{\phi}\sum_{n=-\infty}^{\infty} J_{n}(A) \exp\left\{i \left[\left(\Delta + n \frac{\omega_{p1}+\omega_{p2}}{2}\right)t + \varphi_n^\prime\right]\right\},\\		
		\end{split}
		\label{phase_coupling}
	\end{equation}

\end{widetext}
	where 
\begin{widetext}
	\begin{equation}
		\begin{split}
		C_{\phi} &=g \exp\left\{i\left[\frac{\epsilon_{p1}}{\omega_{p1}}\sin(\phi_{p1})-\frac{\epsilon_{p2}}{\omega_{p2}}\sin(\phi_{p2})\right]\right\},\\
		A^2 &= \left(\frac{\epsilon_{p1}}{\omega_{p1}}\right)^2+\left(\frac{\epsilon_{p2}}{\omega_{p2}}\right)^2
		-2\frac{\epsilon_{p1}\epsilon_{p2}}{\omega_{p1}\omega_{p2}}\cos((\omega_{p1}-\omega_{p2})t+\delta\phi_p), \\
		\varphi &=\frac{\epsilon_{p2}\omega_{p1}-\epsilon_{p1}\omega_{p2} }{\epsilon_{p2}\omega_{p1} + \epsilon_{p1}\omega_{p2}}\cot\left(\frac{\omega_{p1}-\omega_{p2}}{2}t+\frac{\delta\phi_p}{2}\right),\\
		\varphi_n^\prime &= n\left(\frac{\phi_{p1}+\pi + \phi_{p2}}{2} + \varphi\right).\\
		\end{split}
	\end{equation}
\end{widetext}
 We use sum-to-product identities followed by the Jacobi-Anger expansions to simplify the above expression in Eq. \eqref{phase_coupling}, rather than applying the separate Jacobi-Anger expansion \cite{Li2018, Cai2019, Zheng2022, Shi2023, Rosen2024, Jiang2025}. The coupling strength in Eq. \eqref{phase_coupling} becomes time-independent 
 \begin{equation}
 	g_{\text{phase}}^n  = g J_n(A), \label{eq:gphase}
 \end{equation}
 when $\omega_{p} = \omega_{p1} = \omega_{p2}$ and $ \Delta +n\omega_{p} = 0, n \in \mathbb{Z}$, 
  which results in 
  \begin{equation*}
  	\begin{split}
  		A =&
  		\mathrm{sgn}\left[ - \sin\frac{\delta \phi_p}{2}\right]  \\
  		&\times\sqrt{\left(\frac{\epsilon_{p1}}{\omega_{p}} -\frac{\epsilon_{p2}}{\omega_{p}}\cos\delta \phi_p  \right)^2+ \left(\frac{\epsilon_{p2}}{\omega_{p}}\sin\delta \phi_p  \right)^2}.
  	\end{split}
  \end{equation*}


  Additionally, the coupling strength $ g_{\text{phase}}^n $ depends not only on the parametric amplitudes $ \epsilon_{p1},\epsilon_{p2}$ and frequencies $\omega_{p1},\omega_{p2}$, but also on the phases $\phi_{p1}, \phi_{p2}$. This indicates that the parametric relative phase $\delta\phi_p$ can be modulated to adjust the argument $A$ of the $n$th Bessel function, thereby modulating the coupling strength. The range of  $|A|\in \left[ \left| {\epsilon _{p1}}/{\omega _{p1}}-{\epsilon _{p2}}/{\omega _{p2}} \right|,{\epsilon _{p1}}/{\omega _{p1}}+{\epsilon _{p2}}/{\omega _{p2}} \right] $  determines the strength range, which is analogous to interference effects. When the parking flux bias $\bar{\Phi}$ is set at the sweet spot with the parametric amplitude $ \tilde{\Phi} \lesssim \Phi_0/2$, the form of the effective coupling remains the same as in Eq. \eqref{phase_coupling} but with the parametric frequency $\omega_{p1}~(\omega_{p2})$ and phase $\phi_{p1}~(\phi_{p2})$ doubled.

	 \begin{figure}[h]
	\centering
	\includegraphics{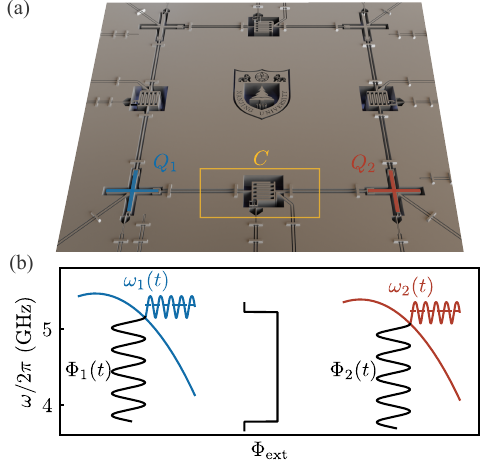}
	\caption{Schematics of the experimental system and parametric phase modulation. (a) False-colored sketch of the superconducting circuit, showing the chip layout with four transmon qubits and four couplers. Qubits, $Q_1$ (blue) and $Q_2$ (red), along with the coupler $C$ (orange), are selected for the experiment. (b) Schematic of applied flux pulses. The dc flux bias and rf flux pulse (black lines) are delivered via dedicated on-chip lines.  A dc flux biases the coupler $C$ to set a desired qubit-qubit coupling strength. Simultaneously, two parametric flux pulses are applied to $Q_1$ and $Q_2$ to induce time-varying qubit frequencies, thereby mediating the phase-controlled coupling. The oscillating blue (red) solid lines illustrate the instantaneous modulated frequencies of $Q_1$ ($Q_2$), while the dashed lines indicate their respective time-averaged frequencies.}
	\label{fig:demo}
	\end{figure}
	
	 \section{Experiment}
	 \subsection{Chip parameters}
The experiment is conducted using a two-qubit system coupled via a tunable coupler in a symmetric  configuration on a superconducting quantum chip. The chip includes four grounded transmon qubits and four floating couplers \cite{Koch2007,Sete2021a}, fabricated using standard lithographic techniques on a high-resistivity silicon substrate \cite{Zheng2022}. The qubits and couplers consist of superconducting quantum interference device (SQUID) loops with symmetric Josephson junctions. The two qubits, $Q_1$ and $Q_2$, and the coupler $C$ are manipulated to perform parametric phase modulation, as shown in Fig. \ref{fig:demo}. The frequencies and anharmonicities of $Q_1~(Q_2, C)$ are $\omega_{1}/2\pi = 5.477~(\omega_{2}/2\pi = 5.401, ~\omega_{c}/2\pi = 5.390)$ GHz and $\alpha_{1}/2\pi = -248 ~(\alpha_{2}/2\pi =-248,~\alpha_{c}/2\pi =-184)$ MHz at the sweet spots, respectively, with coherence times $T_1=16.2 ~(13.9,~12.2)$ $\mu$s and $T_2^\star= 17.3~(20.0, ~4.5)$ $\mu$s. The fixed coupling strengths $g_{1c}/2\pi ~(g_{2c}/2\pi)$ between $Q_1~ (Q_2)$ and $C$ are $115~(78)$ MHz, with a direct coupling strength of $g_{12}/2\pi = 7.5$ MHz between the two qubits. Two parametric flux pulses, $$\Phi_1(t)=\bar{\Phi}_1+\tilde{\Phi}_1\cos(\omega_{p1} t+\phi_{p1})$$ and $$\Phi_2(t)=\bar{\Phi}_2+\tilde{\Phi}_2\cos(\omega_{p2} t+\phi_{p2}),$$ are applied to the two qubits. Additionally, a static dc flux bias $\Phi_c = 0.093\Phi_0$ is used to bias the coupler to achieve an appropriate two-qubit coupling strength $2g/2\pi=21$ MHz (see Appendix~\ref{app:coupler} for details), which depends on the frequencies and fixed coupling strength of this system \cite{Sete2021}.
	 
	 \subsection{Parametric phase modulation}
	A transmon qubit under modulation exhibits various characteristics, requiring precise calibration for parametric phase modulation. When a tunable transmon is modulated, its time-averaged qubit frequency shifts due to the nonlinearity of transmon qubits. To track the frequency excursion, we perform three-tone spectroscopy experiments (see Appendix~\ref{app:spectroscopy} for details).  Notably, the amplitude-frequency response results in greater attenuation at higher parametric frequencies, which can be attributed to the hardware and fridge lines, commonly described by the transfer function (see Appendix~\ref{app:transferfunction} for details). Moreover, a tunable qubit driven by a parametric pulse can bridge the gap between two far-detuned qubits. Both $Q_1$ and $Q_2$ are tunable, and each qubit can  be driven independently by a parametric pulse to induce effective two-qubit coupling (see Appendix~\ref{app:pmcoupling} for details). These factors determine the calibrated operating points and pulse parameters (e.g., $\bar{\Phi}_1,\tilde{\Phi}_1,\omega_{p1},\bar{\Phi}_2,\tilde{\Phi}_2,\omega_{p2} $) for subsequent parametric phase modulation.  
	
	 	Parametric phase modulation is a general method for adjusting coupling strength through phase. Both qubits are simultaneously subjected to two parametric pulses with the same sideband-resonant frequency, and the parametric relative phase modulates the effective two-qubit coupling strength. In the experiment, we demonstrate phase modulation for the first-order sideband couplings, tested at both sweet and off-sweet spots. We prepare the initial state $|10\rangle$ by applying a $\pi$ pulse on $Q_1$, followed by the simultaneous application of two parametric pulses to the qubits with the parameters below. At the sweet spots, we set the qubits' bias to $\bar{\Phi}_1 = \bar{\Phi}_2=0$. To control the first-order sideband coupling, we set $\tilde{\Phi}_1=0.08\Phi_0$, $\tilde{\Phi}_2=0.13\Phi_0$, and $ \omega_{p1}/2\pi =  \omega_{p2}/2\pi = 70.8$ MHz. For the first-order sideband coupling at off-sweet spots, we use $\bar{\Phi}_1 = 0.064\Phi_0$, $\tilde{\Phi}_1 = 0.08\Phi_0$, $\bar{\Phi}_2= 0.1025\Phi_0$, $\tilde{\Phi}_2 = 0.08\Phi_0$, and $\omega_{p1}/2\pi =  \omega_{p2}/2\pi = 178.64$ MHz. The relative parametric phase $\delta\phi_p$ is observed to control the period of population oscillations between the $|10\rangle$ and $|01\rangle$ states. 
	 	
\begin{figure}[h]
	 	\centering
	 	\includegraphics{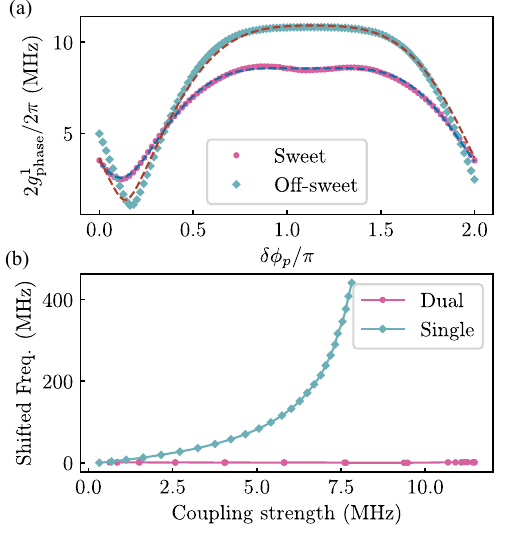}
    \caption{Demonstration of phase-modulated coupling and its suppressed frequency shifts. (a) Phase-modulated coupling strength for the first-order ($n=1$) sideband, $2g_{\mathrm{phase}}^1/2\pi$, achieved with dual parametric pulses and demonstrated at sweet and off-sweet spots. Experimental data points are shown for the sweet spot (fuchsia circles) and the off-sweet spot (teal rhombuses). Corresponding dashed lines represent fits to these datasets using Eq.~\eqref{eq:gphase}, rendered in distinct, high-contrast colors for clarity. The results from both operating conditions demonstrate that the relative parametric phase $\delta \phi_p$ effectively modulates the coupling strength. (b) Comparison of the induced qubit frequency shift required when tuning the coupling strength. Results from our phase-modulation method (fuchsia circles) are contrasted with those from conventional single-pulse amplitude modulation (teal rhombuses), highlighting the significant suppression of frequency shifts with our technique.}
	 	\label{fig:phasemodurabi}
	 \end{figure}

	 	Fig. \ref{fig:phasemodurabi} (a) presents the effective coupling strengths for the first-order ($n=1$) sideband interaction, which are extracted from these oscillations, along with corresponding theoretical fits based on Eq.~\eqref{eq:gphase}. The range of coupling strengths achieved through this phase modulation is substantial, meeting typical requirements for implementing high-fidelity two-qubit gates or performing quantum simulations. To further demonstrate the versatility of our method, we have also applied and characterized this phase-modulation technique for qubits interacting via parametric resonance (i.e., the zeroth-order sideband), with detailed results provided in Appendix~\ref{app:parametricresonance}. A key advantage of our scheme is the suppression of frequency shifts when tuning the first-order coupling strength. This is explicitly demonstrated in Fig.~\ref{fig:phasemodurabi}(b), which numerically compares the frequency shifts induced by our phase-modulation scheme against those from a conventional single-pulse amplitude-modulation scheme. While the conventional method requires large adjustments to the parametric frequency to achieve a desired coupling strength, our scheme induces only negligible shifts across the entire tuning range. Furthermore, our scheme can achieve stronger coupling strengths, extending beyond the practical limits of the conventional amplitude-tuning approach \cite{Didier2018}. Parametric phase modulation is a feasible method, requiring only the adjustment of the time-averaged qubit frequencies to achieve sideband resonance with  dual parametric pulses.

	 In theory, the extremum of the phase-tunable coupling strength occurs at $\delta \phi_{p}=0$ for the first-order sideband coupling. However, two factors cause deviations from the extremum. First, the frequencies of the tunable transmon qubits depend nonlinearly on the external flux, which leads to a nonlinear modulated frequency response and introduces additional phases beyond the linear region. Second, distortions in the actual pulses arise due to limitations of the microwave instruments and fridge lines, which inevitably introduce shifted local phases. 
	
\begin{figure}[h]
 	\centering
 	\includegraphics{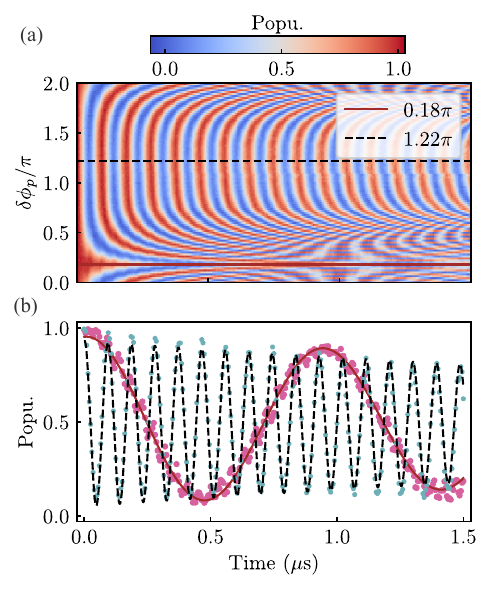}
 	\caption{Modulation of population oscillations between states $|10\rangle$ and $|01\rangle$ by the relative parametric phase $\delta\phi_p$.  (a)~Chevron pattern illustrating the population dynamics as a function of relative parametric phase $\delta\phi_p$ and evolution time. Oscillations at $\delta \phi_p  =0.18\pi$ and $\delta \phi_p  =1.22\pi$ (further detailed in panel (b)) are highlighted within the pattern. (b)~Corresponding population oscillations versus evolution time at $\delta \phi_p  =0.18\pi$ (fuchsia circles, fitted by brown solid line) and $\delta \phi_p  =1.22\pi$ (teal stars, fitted by black dashed line). These traces demonstrate the phase-controlled modulation of the oscillation frequency, and thus the coupling strength.}
 	\label{fig:switch}
 \end{figure}

 	We choose parametric phase modulation using the first-order sideband coupling at the off-sweet spot for an illustration. The population oscillations between $|10\rangle$ and $|01\rangle$ are shown in Fig. \ref{fig:switch}(a), where the relative phase $\delta\phi_p$ modulates the coupling strength $2g_{\text{phase}}^1 $. The coupling strength is largest at $\delta \phi_{p} = 1.22\pi$ and smallest at $\delta \phi_{p} = 0.18\pi$, as shown in Fig. \ref{fig:switch}(b). This demonstrates that the relative parametric phase $\delta \phi_{p}$ can modulate the coupling strength $2g_{\mathrm{phase}}^1$, while keeping the parametric amplitudes and frequencies fixed.
 	
\begin{figure}[h]
		\centering
		\includegraphics{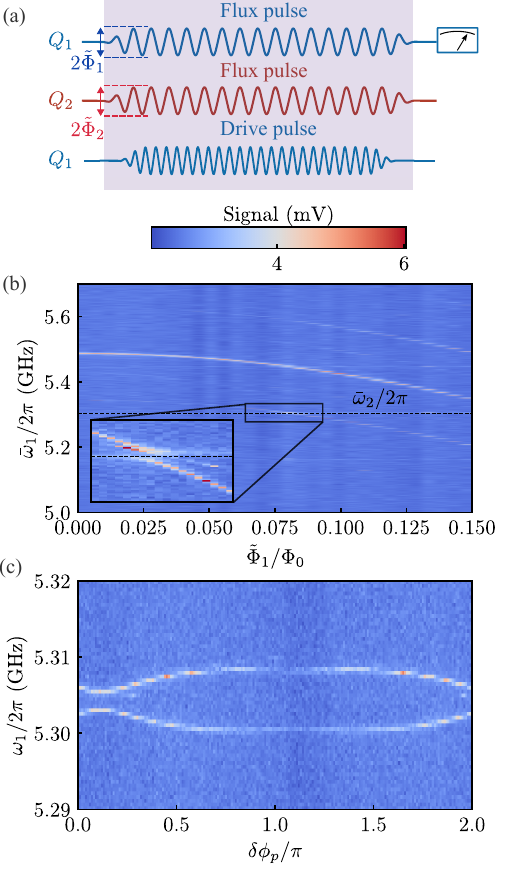}
		\caption{Spectroscopic observation of phase-modulated avoided crossings demonstrating tunable coupling. (a) Pulse sequence for the four-tone spectroscopy used to measure the $Q_1$ spectrum. (b) Measured spectrum of $Q_1$ under dual parametric pulses (on $Q_1$ and $Q_2$) at frequency $\omega_p/2\pi=70.8$ MHz, plotted as a function of varying parametric amplitude $\tilde{\Phi}_1$ on $Q_1$ (with $\tilde{\Phi}_2$ fixed at $0.13\Phi_0$). The inset shows an enlarged view of the avoided crossing corresponding to the first-order ($n=1$) sideband coupling, observed at $\tilde{\Phi}_1=0.08\Phi_0$. (c) The gap of the avoided crossing as a function of the relative parametric phase $\delta\phi_p$ between the dual parametric pulses. This demonstrates that $\delta\phi_p$ directly modulates the gap, i.e., the effective phase-tunable coupling strength $2g_{\text{phase}}^1/2\pi$.}
		\label{fig:specphase}
	\end{figure}

	A spectroscopy experiment is a simple and effective technique for tracking the qubit frequency excursion. It also provides an intuitive way to determine the coupling strength from the gap in the avoided crossing. We conduct the spectroscopy experiment to observe the first-order sideband coupling at off-sweet spots, where the spectrum of $Q_1$ under parametric phase modulation is displayed in Fig. \ref{fig:specphase}. This experiment is similar to the aforementioned three-tone spectroscopy for calibration, but with an additional parametric pulse applied to $Q_2$ at the same frequency called four-tone spectroscopy, as illustrated in Fig. \ref{fig:specphase}(a). An evident avoided crossing, corresponding to the effective coupling, is observed at the relative parametric phase $\delta\phi_p=0$, as seen in Fig. \ref{fig:specphase}(b). The gap of the avoided crossing can be modulated by adjusting the parametric phase $\delta\phi_p$, while keeping other parameters fixed, as demonstrated in Fig. \ref{fig:specphase}(c). This behavior is consistent with the results from population oscillations shown in Fig. \ref{fig:switch}(a). This spectroscopic measurement provides a frequency-domain confirmation of the coupling strength that is consistent with the time-domain population oscillations, offering a comprehensive validation of our phase-modulation technique.

	 	We introduce a new scheme for adjusting coupling strength via phase modulation. This scheme is not limited to two identical modulations, which typically brings two frequency-tunable qubits with fixed coupling to the same frequency \cite{Wang2019, Liu2020, Shi2023}. Combined with the coupler, our scheme extends conventional modulated techniques and enables flexible sideband coupling. Therefore, these parameters can offer fine-grained control over interaction Hamiltonians for advanced quantum computations or simulations. 
	 	
	We explore the region where the flux-to-frequency transduction is nonlinear, i.e., qubits are near to or parked at the dc sweet spot. Compared to the near-linear region, the nonlinear region is harder to calibrate due to large frequency deviations under modulation. However, it holds practical significance due to its lower sensitivity to flux noise, as the dephasing rate depends on the power spectral density of the noise and the gradient of the qubit frequency with respect to flux variations $\partial \omega/ \partial \Phi$ \cite{Ithier2005}. 
	 	
	 \section{Conclusion}
	 In parametric modulation, parametric phases typically do not affect the effective coupling strength. However, we demonstrate that parametric phase modulation, using dual parametric pulses inspired by interference effects, can indeed modulate the coupling strength. Population oscillations and spectroscopy are employed to reveal the impact of phase modulation on the coupling strength. Moreover, we provide a systematic methodology, including spectroscopy, population oscillations, Ramsey fringes, and both Taylor and Fourier expansions, to characterize parametric modulation, even in the presence of strong nonlinearity.
	 
	We explore a wide range of tunable coupling strengths with the experimental parameters commonly used in quantum computation and simulation. We experimentally demonstrate the zeroth- and first-order sideband coupling at both sweet and off-sweet spots. This approach offers a general method for modulating coupling strength, even for higher-order sidebands, which possesses high stability and scalability (see Appendix~\ref{app:stability} for details). Parametric phase modulation thus provides a new and versatile technique for adjusting coupling strength and extends the boundaries of parametric modulation. It is an additional, flexible tool for quantum simulations, and its application to reducing conditional phase or leakage errors in high-fidelity two-qubit gates is a promising avenue for future work.
 
\section*{DATA AVAILABILITY}
The data that support the findings of this paper are openly available at \cite{Ma2025}.

\begin{acknowledgments}
This work was supported by the Innovation Program for Quantum Science and Technology (Grant No. 2021ZD0301702), the
NSFC (Grant No. U21A20436), the NSF of Jiangsu Province (Grants No. BE2021015-1 and No. BK20232002), the Natural Science Foundation of Jiangsu Province (Grant No. BK20233001), and the Natural Science Foundation of Shandong Province (Grant No. ZR2023LZH002).
\end{acknowledgments}

	\appendix

	 \section{Coherence under parametric phase modulation}
\label{app:coherence}
Coherence is central to high-fidelity operations. For qubits under parametric modulation, coherence is influenced by both their intrinsic properties and the surrounding environment. The flux-noise spectra of qubits determine their dephasing rates, while modulation can sometimes increase dephasing due to effects like multiplicative $1/f $ noise. Beyond flux noise, the qubit's environment, particularly two-level system (TLS) defects, is a common source of decoherence. Theoretical work has shown that frequency modulation can help stabilize relaxation rates and mitigate specific dephasing mechanisms arising from TLSs \cite{Matityahu2021, Matityahu2024}. The stability and quality of qubits can also be affected by slow periodic frequency modulation~\cite{Wudarski2024}.

For the flux noise,  there are powerful mitigation strategies. For instance, operating at specific modulation parameters can create ``dynamical sweet spots" that are insensitive to $1/f$ flux noise \cite{Didier2019, Fried2019}, a technique that has been successfully used to demonstrate high-fidelity controlled-Z (CZ) gates \cite{Hong2020}. The dephasing rate $\Gamma_{\phi,1/f}$ under modulation can be described as 
\begin{equation}
	\Gamma_{\phi,1/f} = \lambda \left|\frac{\partial \bar{\omega}}{\partial \tilde{\Phi}}\right | A_{\rm{ac},1/f},
\end{equation}
where $ \lambda$ and $A_{\rm{ac},1/f}$ represent the noise parameter and amplitude due to $1/f$ noise  \cite{Didier2019}. This concept can be extended. For example, two-tone parametric modulation can create a continuum of dynamical sweet spots, expanding the range of flux-noise-robust operating frequencies \cite{Didier2019a, Valery2022}. The benefits of parametric drives have also been explored for enhancing coherence in bichromatically driven Floquet qubits \cite{BrisenoColunga2025}. Parametric modulation can also be applied to fluxonium qubits to protect qubits from $1/f $ noise and enhance coherence times \cite{Huang2021, Mundada2020}. Indeed, parametric modulation can be viewed as a continuous version of the dynamical decoupling scheme to realize high-fidelity CZ gates which can greatly reduce qubit dephasing \cite{Ma2024}.

Coherence degradation is a known challenge for some hardware-based tunable coupler schemes \cite{Stehlik2021, Li2024}. Such degradation often occurs when the system is tuned into a highly hybridized regime, where the coherence of the qubits can be limited by that of the coupler which can be explained using a two-spin Hamiltonian \cite{Barends2013} or by the resulting dressed states becoming more sensitive to flux noise. The effective decoherence rates in such systems can be modeled by the participation ratios of the uncoupled modes \cite{Marxer2023}.

In our experiment, we have designed our scheme to specifically enhance coherence. We choose two qubits with a coupler in our chip, and other irrelevant couplers are biased to idle points with zero couplings. The chosen hardware coupler is biased to a fixed operating point that maintains a sufficient effective coupling strength while ensuring the qubits are only weakly hybridized with it. Crucially, our working points for the qubits are at or near to their dc sweet spots, leveraging their inherent insensitivity to flux noise \cite{Ithier2005}. Consequently, the dual parametric drives, which are simultaneously applied to the two adjacent qubits, are designed to inherit the robustness of parametric modulation against pulse distortions and noise. The introduction of a second, synchronous drive primarily adds a new degree of control---the relative phase---without fundamentally altering the underlying noise-protection mechanisms.

\section{Spectroscopy of parametrically modulated qubits}
\label{app:spectroscopy}

\begin{figure*}[htbp] 
	\centering
	\includegraphics{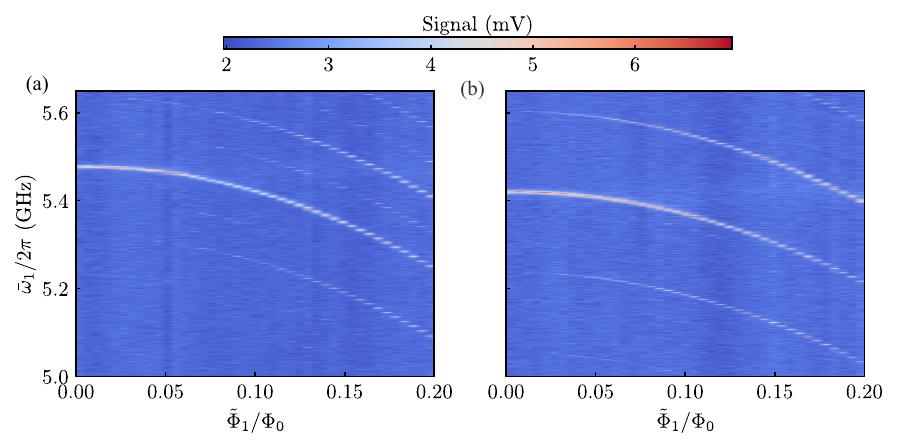}
	\caption{Three-tone spectroscopy of a parametrically modulated qubit, $Q_1$. (a) Spectrum of $Q_1$ at its sweet spot.  A parametric pulse at a frequency of $\omega_{p1}/2\pi=79.2$ MHz is applied with varying amplitude $\tilde{\Phi}_1$. Only even-order sidebands are prominent, while odd-order sidebands are strongly suppressed. (b) Spectrum of $Q_1$ at an off-sweet spot ($\bar{\Phi}_1 = 0.064\Phi_0$). A parametric pulse at $\omega_{p1}/2\pi = 181.2$ MHz is applied with varying amplitude $\tilde{\Phi}_1$. Sidebands of all integer orders are observed. }
	\label{fig:specvsac}
\end{figure*}

To demonstrate the effect of parametric flux pulses on qubits, we implement a three-tone spectroscopy experiment, using two microwave tones and a parametric flux tone, to capture the average response of qubit frequencies under parametric modulation \cite{Li2013, Rosen2024}. Unlike the two-tone spectroscopy experiment, an additional parametric flux pulse is applied to the qubit. We perform spectroscopy experiments on $Q_1$ at sweet spot, $\bar{\Phi}_1=0$, and off-sweet spot, $\bar{\Phi}_1=0.064\Phi_0$, with fixed parametric frequencies of $\omega_{p1}/2\pi=79.2$ MHz and $181.2$ MHz, respectively, as shown in Fig. \ref{fig:specvsac}. We observe that the time-averaged qubit frequencies decrease as the parametric amplitudes increase with the fixed parametric frequency, which arises from the nonlinearity of the transmon qubit to the external flux. When the parametric amplitude approaches zero, the spectrum shows only a peak, representing the dc component of the qubit’s modulated frequency. However, as the parametric amplitude increases, harmonic frequency peaks appear, corresponding to the sideband frequencies of the modulated qubit. At off-sweet spots, these sidebands follow the pattern $\bar{\omega}_1 + n\omega_{p1}$,  and at sweet spots, they follow $\bar{\omega}_1 + 2n\omega_{p1}$, where $n \in \mathbb{Z}$. 

This relationship between the time-averaged frequencies and parametric amplitudes provides insights into the parametric sideband-resonant conditions of two qubits and facilitates further calibration. At sweet spots, the first-order sideband frequencies $\bar{\omega}_1 \pm \omega_{p1}$ theoretically disappear, which contrasts with the behavior at off-sweet spots. This is because modulated qubits undergo double cycles compared to the parametric flux pulses at sweet spots \cite{Caldwell2018}. As shown in Fig. \ref{fig:specvsac} (a), the first-order sidebands are nearly invisible because of the small flux deviations at sweet spots. The parametric amplitude $\tilde{\Phi}_{1}$ (related to the frequency excursion $\epsilon_{p1}$) determines the effective coupling strength $g_{\mathrm{eff}}^n$ of different sidebands via Bessel functions, as described in Eq. \eqref{effcoupling}. Consequently, variations in  $\tilde{\Phi}_{1}$ directly impact the observed prominence and intensity of these sidebands in the spectrum shown in Fig. \ref{fig:specvsac}.  An interesting phenomenon is that the linewidth of the modulated qubit broadens with larger amplitudes due to power broadening \cite{Li2013}.

The spectroscopy experiment provides a straightforward and detailed method to observe the spectrum of qubits under parametric modulation. The spectrum reveals frequency excursions, potential avoided crossings, and even possible errors in the frequency domain. The simplicity and richness of this experimental approach make it widely applicable in this study.

	 \section{Numerical comparison between the Taylor expansion and the Fourier series}
\label{app:taylor}
Transmon qubits can be accurately modeled as a combination of charge and Josephson energies, using either a cosine or harmonic potential \cite{Koch2007, Willsch2024}. However, for tunable transmon qubits with time-varying external flux, subtle differences arise due to capacitance ratios between branches and realistic circuit geometries \cite{You2019, Rajmohan2022, Riwar2022, Petrescu2023}. To simplify, we analyze the flux-tunable transmon using a $25$th-order analytical equation in a positive real number $\xi$ (related to the zero-point fluctuations), with its numerical accuracy validated against the Fourier series \cite{Didier2018}. The analytical equation, which depends on external flux, is essential for fitting experimental data and extracting qubit features.

The dependence of qubit frequencies on external flux is nonlinear. In the main text, we use the Fourier series to describe the frequency response under parametric modulation. Alternatively, we can also employ a Taylor expansion to capture the qubit frequency behavior. While the Fourier series provides a fast, accurate method for estimating harmonics, it is challenging to measure experimentally. In contrast, the Taylor expansion, which consists of derivatives of qubit frequency with respect to flux, can be directly derived from the spectrum and is experimentally feasible for obtaining harmonics and predicting parametric coupling between two qubits \cite{McKay2016,Han2020}.

We show the Taylor expansion of the qubit frequency for a flux pulse parameterized as $\Phi(t) = \bar{\Phi} + \tilde{\Phi} \cos(\omega_p t)$ (assuming parametric phase $\phi_p=0$). The qubit frequency can be expanded as 
\begin{equation}
	\omega(t) = \omega(\bar{\Phi}) + \sum_{n=1}^\infty \frac{1}{n!} \left.\frac{\partial^n \omega}{\partial \Phi^n}\right|_{\bar{\Phi}} \left[ \tilde{\Phi} \cos(\omega_p t) \right]^n.
	\label{eq:taylor}
\end{equation}
Using the power-reduction formulas, we can rewrite the Taylor expansion \eqref{eq:taylor} as 
\begin{equation}
	\begin{split}
		\omega(t) &=\left[\omega(\bar{\Phi}) + \sum_{n=1}^\infty \frac{\tilde{\Phi}^n}{2^nn!} \left.\frac{\partial^n \omega}{\partial \Phi^n}\right|_{\bar{\Phi}}\delta_{{n\bmod2},0}\binom{n}{\lfloor{n}/{2}\rfloor}\right] \\
		&+  \sum_{n=1}^\infty \frac{2\tilde{\Phi}^n}{2^{n}n!} \left.\frac{\partial^n \omega}{\partial \Phi^n}\right|_{\bar{\Phi}}\sum_{k=0}^{\lfloor{(n-1)}/{2}\rfloor}\binom{n}{k} \cos ((n-2 k) \omega_p t).
	\end{split}
	\label{eq:taylor1}
\end{equation}
As the expansion order $n$ increases, more harmonics appear, improving accuracy. We choose typical parameters $\bar{\Phi} = 0,~\tilde{\Phi} = 0.4\Phi_0 $  at sweet spots and $\bar{\Phi} = 0.15\Phi_0,~ \tilde{\Phi} = 0.3\Phi_0 $ at off-sweet spots as examples.

\begin{figure}[htbp] 
	\centering
	\includegraphics{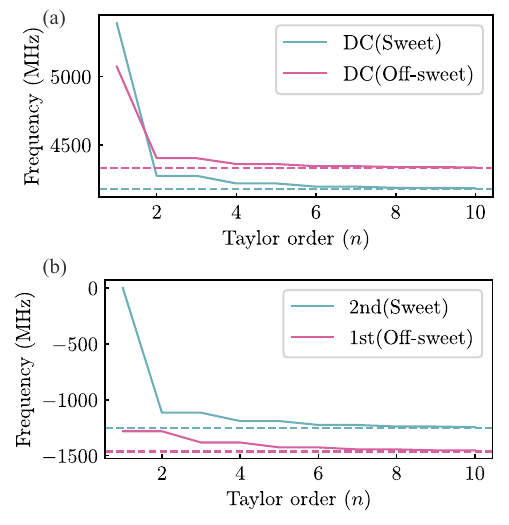}
	\caption{Comparison of Taylor expansion and Fourier series for calculating components of the parametrically modulated qubit frequency. (a) The dc frequency shift of the qubit under parametric modulation, shown as a function of the expansion order $n$. Results are presented for modulation at the sweet spot ($\bar{\Phi}=0, \tilde{\Phi}=0.4\Phi_0$; teal lines) and an off-sweet spot ($\bar{\Phi}=0.15\Phi_0, \tilde{\Phi}=0.3\Phi_0$; fuchsia lines). Solid lines represent calculations from Taylor expansion, while dashed lines show the corresponding Fourier-series components. (b) Amplitude of the dominant ac harmonic component of the modulated qubit frequency versus expansion order $n$. For the sweet spot (teal lines), this is the second harmonic (frequency $2\omega_p$ component); for the off-sweet spot (fuchsia lines), it is the first harmonic (frequency $\omega_p$ component). Taylor-expansion results (solid lines) are shown approaching the Fourier-series values (dashed lines) as $n$ increases. All calculations use transmon qubit parameters $E_C/2\pi=240$ MHz and $E_{J1}/2\pi=E_{J2}/2\pi=8.286$ GHz. }
	\label{fig:functionappro}
\end{figure}

As shown in Fig. \ref{fig:functionappro} (a), the dc shifts depend on the even-order terms of the Taylor expansion. The main shift can be approximated by the second-order ($n=2$) term. As the Taylor order $n$ increases, the dc component approaches the zeroth order of the Fourier series, corresponding to the time-averaged frequency for one period. The dc component results from the nonlinearity in transmon qubits and can also be derived using the charge and flux operators, as shown in Ref. \cite{Rosen2024}, rather than the analytical equation provided above.

As shown in Fig. \ref{fig:functionappro} (b), we respectively estimate the second harmonic at the sweet spot and the first harmonic at the off-sweet spot, which are used to theoretically determine the parametric coupling strength at weak parametric modulation amplitudes. The main harmonic components are obtained by considering the Taylor expansion at order $n=2$ (the sweet spot) and at order $n=3$ (the off-sweet spot). At sweet spots, the qubit frequency oscillates only at even harmonics of the parametric frequency since all odd-order derivatives vanish. No significant difference is observed for Taylor-expansion orders $n>6$ at sweet and off-sweet spots. 

\begin{figure}[htbp] 
	\centering
	\includegraphics{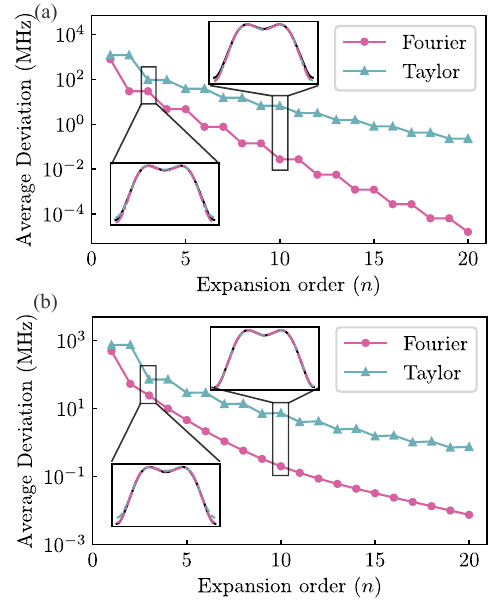}
	\caption{Average deviations of Taylor-expansion and Fourier-series approximations from an analytical solution for the parametrically modulated qubit frequency $\omega(t)$. (a) Average deviation over one modulation period from the analytical solution for calculations at the sweet spot, plotted as a function of increasing expansion order $n$. (b) Corresponding average deviation for calculations at the off-sweet spot. In both panels (a) and (b), results for the Fourier series are shown as solid lines with circle markers, while results for the Taylor expansion are shown as solid lines with triangle-up markers. Insets: Time evolution of the modulated qubit frequency, $\omega(t)$, over one period, comparing the analytical solution (black solid line) with approximations from Fourier series  (fuchsia dashed line) and Taylor expansion (teal dotted line) at expansion orders $n=3$ and $n=10$. All physical and modulation parameters are identical to those used in Fig. \ref{fig:functionappro}.} 
	\label{fig:functionapproerror}
\end{figure}

\begin{figure}[htbp] 
	\centering
	\includegraphics{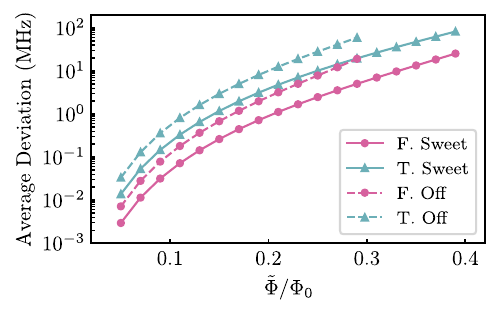}
	\caption{Average deviations of Taylor-expansion and Fourier-series approximations (at a fixed expansion order of $n=3$) from an analytical solution for the parametrically modulated qubit frequency, shown as the parametric amplitude $\tilde{\Phi}$ increases. Solid and dashed lines represent calculations for the qubit at its sweet and off-sweet spots, respectively. Fuchsia lines correspond to the Fourier series approximation, and teal lines correspond to the Taylor expansion approximation. All other physical and modulation parameters are identical to those specified in Fig. \ref{fig:functionappro}.}
	\label{fig:errorvsamp}
\end{figure}

The time-dependent frequency of qubits under parametric modulation can be approximated using both the Taylor expansion and the Fourier series. The average deviations over a period between these methods and the analytical equation decrease as the expansion order $n$ increases at sweet and off-sweet spots, as shown in Fig. \ref{fig:functionapproerror}. The insets show the time evolution of qubit frequencies over one period at orders $n=3$ and $n=10$. At $n=3$, the Taylor expansion slightly deviates, while no visible deviations occur at higher orders. At sweet spots, the Fourier spectrum exhibits a stepped shape due to the vanishing of odd Fourier coefficients \cite{Didier2018}. Average deviations are lower at sweet spots compared to off-sweet spots, due to the symmetry of flux-to-frequency transduction. Both methods provide good approximations of qubit time-dependent frequencies and coupling-strength calculations for weak parametric amplitudes, but the accuracy decreases as the parametric amplitude increases, as shown in Fig. \ref{fig:errorvsamp}. At sweet spots, both methods yield lower deviations due to the symmetry of flux-to-frequency transduction, which eliminates odd harmonics.

	 \section{Tunable coupling strength via a coupler}
\label{app:coupler}
\begin{figure}[htbp] 
	\centering
	\includegraphics{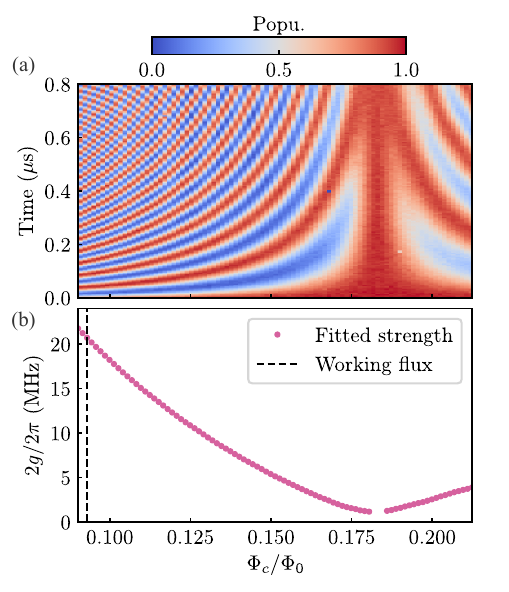}
	\caption{Qubit-qubit coupling strength $2g$ as a function of the coupler flux bias $\Phi_c$. (a) Measured population of the state $|10\rangle$, exhibiting oscillations as a function of evolution time and $\Phi_c$. (b) Extracted coupling strength $2g$, obtained by fitting the population oscillations from panel (a), plotted against $\Phi_c$. The vertical black dashed line indicates the selected working coupler flux, $\Phi_c = 0.093\Phi_0$, with a coupling strength $2g/2\pi=21$ MHz used in the experiment. }
	\label{fig:strengthvsflux}
\end{figure}

In the two-qubit system featuring a coupler, the effective qubit-qubit coupling strength can be controlled by biasing the coupler flux. We measure the qubit-qubit coupling strength, $2g$, as a function of the coupler flux. Initially, we prepare the qubit in the state $|10\rangle$ using a $\pi$ pulse, then adjust the coupler flux to measure population oscillations between the states $|10\rangle$ and  $|01\rangle$ after bringing the two qubits into resonance, as shown in Fig. \ref{fig:strengthvsflux}(a). These oscillations are fitted using a cosine decay model to extract the coupling strength $2g$, as shown in Fig. \ref{fig:strengthvsflux}(b).  In this experiment, we choose a working coupler flux $\Phi_c = 0.093\Phi_0$, which maintains a coupling strength $2g/2\pi=21$ MHz. The idle coupler flux is achieved by biasing the coupler to zero coupling strength, effectively decoupling the qubits from coupler-mediated interactions.

	 \section{Flux pulse transfer function}
	 \label{app:transferfunction}
	 
	 \begin{figure}[ht] 
	 	\centering
	 	\includegraphics{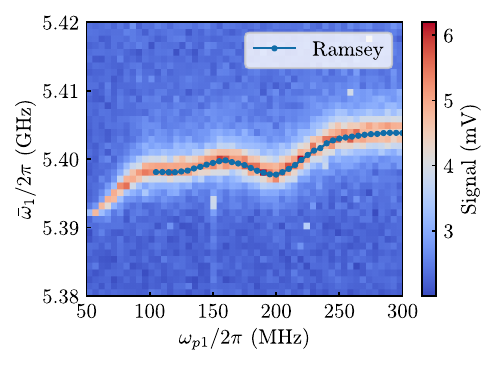}
	 	\caption{Characterization of the flux pulse transfer function for $Q_1$ using a Ramsey-like experiment and three-tone spectroscopy, both performed with a nominal applied parametric amplitude of $\tilde{\Phi}_1=0.065\Phi_0$. The transfer function (effective qubit frequency excursion versus frequency $\omega_{p1}/2\pi$) is determined from the period of Ramsey fringes in the Ramsey-like method. Both methods yield consistent results for parametric frequencies $\omega_{p1}/2\pi > 100$ MHz. However, at lower parametric frequencies, the Ramsey-like method becomes less reliable; the reduced attenuation in the flux line results in a larger actual modulation amplitude at the qubit, inducing significant qubit frequency excursions. These large excursions cause the Ramsey fringes to oscillate too rapidly (i.e., with a very short period) to be accurately resolved with available microwave instrument sampling. Three-tone spectroscopy, in contrast, remains effective for characterizing the frequency excursion in this low-frequency regime.}
	 	\label{fig:pulsetransferfunction}
	 \end{figure}
	The parametric pulse offers several advantages over the unipolar flux pulse \cite{Rol2019}, particularly in terms of flux pulse distortions. While a conventional unipolar flux pulse can have significant power in multiple frequency components after a Fourier transform, a parametric flux pulse has a single frequency. This feature helps avoid distortions arising from the collective response of different frequencies, considering microwave devices, electrical components, and on-chip response \cite{Rol2020}.
	
	 To measure the transfer function \cite{Sete2021, Valery2022}, which describes the dependence of actual parametric amplitudes on parametric frequencies, we use two characterization methods: a Ramsey-like experiment and three-tone spectroscopy. The Ramsey-like pulse sequence adds an additional parametric flux pulse between two $\frac{\pi}{2}$ pulses, compared to the conventional Ramsey experiment. This method is quick and provides a means to characterize the time-averaged qubit frequency excursion at different parametric frequencies, as shown in Fig. \ref{fig:pulsetransferfunction}. The transfer function of $Q_1$ depends on chip design, specific microwave devices, and circuit elements. At the same parametric amplitude, $\tilde{\Phi}_1=0.065\Phi_0$, the time-averaged qubit frequency decreases as the parametric frequency decreases, suggesting that the actual parametric amplitude at the qubit increases, as observed in Fig. \ref{fig:pulsetransferfunction}. A notable feature of the transfer function is the increased attenuation at higher parametric frequencies. However, as the parametric frequency approaches zero (i.e., for a static dc flux bias), the frequency excursion may become too large to be captured by the Ramsey-like experiment. Therefore, we use three-tone spectroscopy as a supplementary tool to capture the full qubit frequency excursion. The results from both methods are consistent when the parametric frequency $\omega_{p1}/2\pi$ exceeds $100$ MHz. The transfer function provides the effective modulation amplitude at the qubit as a function of parametric frequencies, aiding in the selection of optimal modulation parameters. 
 
	 \section{Coupling strength of different parametric amplitudes}
	 \label{app:pmcoupling}

\begin{figure}[!h] 
	\centering
	\includegraphics{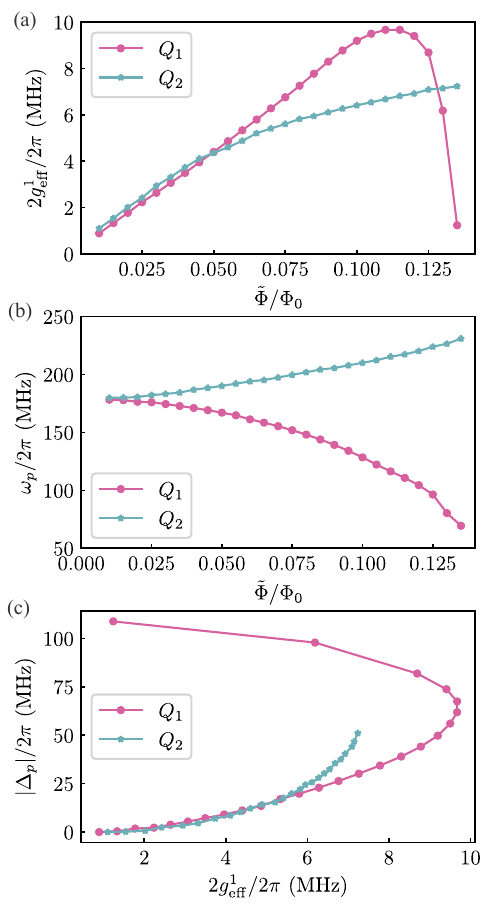}
	\caption{Dependence of first-order ($n=1$) sideband coupling on the parametric drive amplitude $\tilde{\Phi}$. 
		The data are for scenarios where either $Q_1$ or $Q_2$ is independently driven, while the other qubit is maintained at a static dc flux bias. 
		(a) Effective coupling strength $2g_{\mathrm{eff}}^1/2\pi$ versus parametric amplitude $\tilde{\Phi}$, extracted from the period of population oscillations between the $|10\rangle$ and $|01\rangle$ states. 
		(b) Parametric frequency $\omega_p/2\pi$ required to maintain the sideband resonance condition, plotted against $\tilde{\Phi}$. 
		(c) The resulting resonant frequency deviation, $|\Delta_p|=|\omega_p(\tilde{\Phi})-\omega_p(0)|$, plotted against the achieved effective coupling strength $2g_{\mathrm{eff}}^1/2\pi$. Panel (c) highlights the significant frequency adjustment required when tuning the coupling via amplitude.}
	\label{fig:couplingvsac}
\end{figure}
	To demonstrate the parametric coupling strengths as parametric amplitudes increase, we focus on the first-order sideband resonance of two qubits at off-sweet spots. In the linear region, where the qubit frequency versus external flux is nearly linear, the relation between parametric coupling strengths and amplitudes follows the first-order Bessel function of the first kind \cite{Chu2020}. However, due to the nonlinearity of transmon qubits, the first-order sideband resonant frequency deviates as the parametric amplitude increases. Additionally, the qubit frequency excursions during modulation, $\epsilon_p$, become increasingly unpredictable (see Appendix \ref{app:taylor} for details).
 
	We bias two qubits at off-sweet spots, with $\bar{\Phi}_1 = 0.119\Phi_0$ and $\bar{\Phi}_2= 0.1025\Phi_0$, to measure the effective coupling strength as a function of the parametric amplitude $\tilde{\Phi}$. As the amplitude increases, the optimal parametric flux frequency for achieving first-order sideband resonance can shift, which can be predicted using a three-tone spectroscopy experiment. In the experiment, we first prepare the $|10\rangle$ state and then impose a parametric flux pulse on $Q_1$ ($Q_2$) while keeping $Q_2$ ($Q_1$) at a static dc flux bias. We measure the population oscillations of the $|10\rangle$ state, and from the oscillations versus parametric frequencies, we can extract both the corresponding effective coupling strengths $2g_{\mathrm{eff}}^1$ [shown in Fig. \ref{fig:couplingvsac}(a)] and the parametric-resonance frequencies $\omega_p$ [shown in Fig. \ref{fig:couplingvsac}(b)]. This amplitude-based tuning method reveals a significant practical challenge: the nonlinearity of the qubit's frequency response to external flux causes the required resonant drive frequency, $\omega_p$, to shift substantially as the parametric amplitude $\tilde{\Phi}$ is varied. As illustrated in Fig. \ref{fig:couplingvsac}(c), this frequency deviation, defined as $|\Delta_p|=|\omega_p(\tilde{\Phi})-\omega_p(0)|$, can exceed $100$ MHz to achieve a desired coupling strength. Such a large covariation necessitates cumbersome, multiparameter calibration routines.  In stark contrast, our phase-modulation scheme provides a decoupled control knob, allowing the coupling strength to be tuned without inducing these parasitic frequency shifts, thereby greatly simplifying the calibration process. 
 
 By combining the transfer function and precalibrated three-tone spectroscopy, we can effectively explore parametric coupling under large frequency deviations, where the flux-to-frequency transduction is highly nonlinear.

	 \section{Parametric-resonance phase modulation}
	 \label{app:parametricresonance}

\begin{figure}[h]
	\centering
	\includegraphics{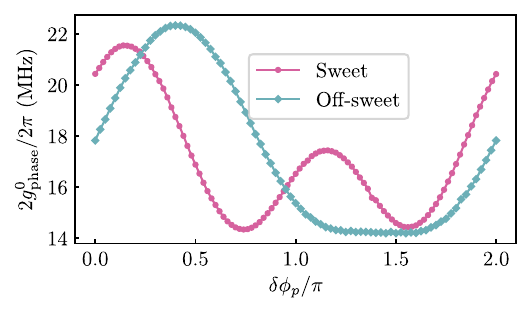}
	\caption{Modulation of the zeroth-order ($n=0$) sideband coupling strength (i.e., parametric resonance), $2g_{\mathrm{phase}}^0/2\pi$, via the relative parametric phase $\delta\phi_p$ of dual parametric pulses. Data points show the coupling strength measured at sweet (fuchsia circles) and off-sweet (teal rhombuses) spots of the qubits. The results demonstrate effective modulation of the direct resonant coupling by $\delta\phi_p$ in both operational regimes.}
	\label{fig:phasemodurabi0th}
\end{figure}

Parametric resonance is a technique for realizing fast two-qubit entangling gates and mitigating issues such as frequency collisions in superconducting qubit systems \cite{Sete2021}. The phase-modulation approach detailed in the main text can also be effectively applied to these parametric-resonance interactions (i.e., the zeroth-order ($n=0$) sideband coupling), and this application is demonstrated in Fig.~\ref{fig:phasemodurabi0th}.

The experimental procedure for achieving phase-modulated coupling via parametric resonance is analogous to that described for the first-order sideband coupling in the main text. The primary distinction lies in the activation condition: for zeroth-order sideband coupling (parametric resonance), the interaction is engaged by tuning the two parametrically modulated qubits such that their time-averaged frequencies become resonant (i.e., $\bar{\omega}_1 = \bar{\omega}_2$). This contrasts with first-order sideband couplings, where resonance is achieved when the effective detuning matches a nonzero integer multiple of the parametric flux frequency (e.g., $\Delta + \omega_p = 0$).

The experimental parameters employed for demonstrating phase modulation of the zeroth-order sideband coupling are as follows. At sweet spots, we use the following parameters: $\tilde{\Phi}_1=0.174\Phi_0$, $\tilde{\Phi}_2=0.1\Phi_0$, and $\omega_{p1}/2\pi = \omega_{p2}/2\pi = 110$ MHz. At off-sweet spots, we adjust the bias and modulation parameters as follows: $\bar{\Phi}_1 = 0.119\Phi_0$, $\tilde{\Phi}_1 = 0.1235\Phi_0$, $\bar{\Phi}_2= 0.1025\Phi_0$, $\tilde{\Phi}_2 = 0.05\Phi_0$, and $\omega_{p1}/2\pi =  \omega_{p2}/2\pi = 290$ MHz.

	 \section{Stability and scalability of parametric phase modulation}
		\label{app:stability}
		
		In this paper, we have defined the main variable for the relative phase between the two drives as $\delta\phi_p$. Fluctuations in this quantity will translate into fluctuations in the coupling strength, $\delta g_{\text{phase}}^n$. To avoid notational ambiguity, we will denote a small, random fluctuation (i.e., jitter) in our relative phase variable as $ j_\phi$.
		
		Fluctuations of absolute parametric phases usually arise from electronic devices, lines, and the thermal and electromagnetic environment. The dominant source of these fluctuations is typically the relative phase jitter between the output channels of the arbitrary waveform generator. For the modern high-performance AWGs used in quantum control, a typical timing jitter of $\sim 5$ ns for a $100-$ MHz parametric drive results in a phase fluctuation of $ j_\phi\sim0.003$ rad \cite{Yang2022}.
		
		We have performed a calculation based on our theoretical model to estimate the impact of this jitter. The fluctuation in coupling strength can be approximated as $\delta g_{\text {phase }}^n\approx|{dg_{\text {phase }}^n}/{d\left(\delta\phi_p\right)}|j_\phi$. Using the chain rule and the properties of Bessel functions, the sensitivity is given by

		 \begin{equation}
			\begin{split}
			\left|\frac{d g_{\text {phase }}^{n}}{d\left(\delta \phi_{p}\right)}\right|&=\left|\frac{d\left(g J_{n}(A)\right)}{d A} \frac{d A}{d\left(\delta \phi_{p}\right)}\right|\\
			&=\left|\frac{g}{2}\left[J_{n-1}(A)-J_{n+1}(A)\right] \frac{d A}{d\left(\delta \phi_{p}\right)}\right|\\
			&=\left|\frac{g}{2}\left[J_{n-1}(A)-J_{n+1}(A)\right] \frac{\epsilon_{p 1} \epsilon_{p 2}}{A \omega_{p}^{2}} \sin \left(\delta \phi_{p}\right)\right|,
			\end{split}
		\end{equation}
 		where $A$ is the phase-dependent argument of the Bessel function. We evaluated this expression at the most sensitive point of our experimental data in Fig. \ref{fig:phasemodurabi} (a) yielding a sensitivity of $|{dg_{\text {phase }}^n}/{d\left(\delta\phi_p\right)}|\approx22$ MHz/rad. The resulting fluctuation in coupling strength is therefore $ \delta g_{\text {phase }}^n\approx0.066$  MHz. Given that our demonstrated coupling strengths are on the order of several megahertz, this corresponds to a relative fluctuation of less than $1\%$ at the most sensitive point. This level of fluctuation is negligible for our experiments and does not require frequent recalibration.

		We utilize a two-qubit system with a tunable coupler to demonstrate our scheme and these qubits are also tunable which is common in large-scale quantum simulation and computation. Regarding wiring complexity, our scheme is highly scalable as it requires no additional physical hardware elements. The dual parametric drives are delivered through the standard on-chip flux bias lines that are already present for individual qubit control in state-of-the-art multiqubit processors. Therefore, it introduces no new hardware overhead.
		
		Regarding crosstalk, we agree that flux crosstalk is a key challenge in scaling superconducting processors. In our scheme, since the parametric drives are applied locally via these individual flux lines, the crosstalk considerations are fundamentally the same as those in any multiqubit system employing simultaneous flux control. We therefore expect that established dc and ac flux crosstalk mitigation techniques, such as those described in Ref. \cite{Abrams2020}, can be directly applied to characterize and suppress these effects. 
		
		In summary, we believe our technique is well suited for practical applications in larger systems because it adds a powerful layer of control without introducing new hardware or wiring complexity and is compatible with existing solutions for crosstalk management. We also note that it may even offer new strategies for its mitigation.


%

\end{document}